\documentclass[prb,twocolumn,superscriptaddress]{revtex4-1}
\usepackage{dcolumn}
\usepackage{bm}
\usepackage{graphicx}
\usepackage{amsmath}
\usepackage{amssymb}

\begin{document}
\title{Hybridization gap and Fano resonance in SmB${_6}$}
\author{S. R\"o{\ss}ler}
\affiliation{Max Planck Institute for Chemical Physics of Solids,
N\"othnitzer Str. 40, 01187 Dresden, Germany}
\author{Tae-Hwan Jang}
\affiliation{Max Planck Institute for Chemical Physics of Solids,
N\"othnitzer Str. 40, 01187 Dresden, Germany}
\author{D. J. Kim} \affiliation{University of California,
Irvine, California 92697, USA}
\author{L. H. Tjeng}
\affiliation{Max Planck Institute for Chemical Physics of Solids,
N\"othnitzer Str. 40, 01187 Dresden, Germany}
\author{Z. Fisk}
\affiliation{University of California, Irvine, California 92697,
USA}
\author{F. Steglich}
\affiliation{Max Planck Institute for Chemical Physics of Solids,
N\"othnitzer Str. 40, 01187 Dresden, Germany}
\author{S. Wirth}
\affiliation{Max Planck Institute for Chemical Physics of Solids,
N\"othnitzer Str. 40, 01187 Dresden, Germany}
\date{\today}

\begin{abstract}
We present results of Scanning Tunneling Microscopy and
Spectroscopy (STS) measurements on the ``Kondo insulator''
SmB$_6$. The vast majority of surface areas investigated was
reconstructed but, infrequently, also patches of varying size of
non-reconstructed, Sm- or B-terminated surfaces were found. On the
smallest patches, clear indications for the hybridization gap and
inter-multiplet transitions were observed. On non-reconstructed
surface areas large enough for coherent co-tunneling we were able
to observe clear-cut Fano resonances. Our locally resolved STS
indicated considerable finite conductance on all surfaces
independent of their structure.
\end{abstract} \maketitle

Materials with strong electron correlations continue to draw
enormous attention not only because they may give rise to
fundamentally new states of matter or new phenomena but also due
to the hope for advanced technological applications. Heavy fermion
(HF) materials, {\it i.e.} intermetallics of certain rare earths
(REs) like Ce, Sm and Yb, are model systems to study strong
electronic correlations \cite{gre91}. Here, the RE-derived
localized $4f$ states are covalently mixed with the
conduction-band states and thus, acquire a finite lifetime. The
associated decay rate in relation to the energy of the localized
4$f$ state corresponds to the valency. In a Sm-based HF system the
valence lies between 3+ ($4f^5$) and 2+ ($4f^6$) which implies a
considerable amount of charge fluctuations. This is usually
referred to as intermediate valence \cite{var76}. In addition to
the above-mentioned mixing of $4f$ states and the conduction band,
which is well described within the framework of one-electron
models, a many-body interaction is operating between the $4f$ and
conduction electrons. This ``Kondo effect'' \cite{kon64}
eventually leads to a screening of the local moments as a result
of particle-hole excitations which are manifested by a narrow
Abrikosov-Suhl, or Kondo, resonance at $E_{\rm F}$ the width of
which is given by the single-ion Kondo temperature $T_{\rm K}$.

Due to the periodic arrangement of REs in a HF intermetallic, the
Kondo resonances couple to a weakly dispersive HF or ``coherent
4$f$-'' band resulting in a heavy Fermi liquid state well below
$T_{\rm K}$. The band interaction between the renormalized 4$f$
and the conduction band generates a so-called hybridization gap
which opens at around $T_{\rm K}$. Under certain conditions,
$E_{\rm F}$ may reside inside this gap characterizing a so-called
Kondo insulator \cite{aep92}.

SmB$_6$ is such a Kondo insulator with a valence $\nu \sim 2.6$
\cite{vai65}, $\nu$ being slightly temperature dependent
\cite{tar80,mit09}. A sharp decrease in the density of states
(DOS) at $E_{F}$ is most dramatically seen in the rise of the
resistance over more than three orders of magnitude below $\sim$40
K \cite{all79,mol82}. Many other measurements, including far
infrared absorption \cite{mol82} and NMR \cite{pen81}, are
consistent with the opening of an activation gap of about 3 meV.
The gradual opening of a hybridization gap of about 18 meV below
$\sim$100 K and signatures of inter-ion correlations below
$\sim$30 K were recently reported based on point-contact
\cite{zha13} and angle-resolved photoemission spectroscopy (ARPES)
\cite{nxu13,neu13,fra13}. However, the gap being caused by
hybridization is discussed controversially. For instance, if the
hybridization is altered by the application of pressure a
continuous change of the gap is expected \cite{bei83,mos85}, while
more recent experiments show a discontinuous change \cite{coo95}.

Intriguingly, unlike in an insulator, the resistance of SmB$_6$
saturates below about 4 K \cite{men69}. Recently, it was proposed
\cite{dze10,tak11} that SmB$_6$ may be a strong three-dimensional
topological insulator where the topologically non-trivial
insulating state is produced by the hybridization between
conduction band and coherent 4$f$ band.
This proposal sparked a flurry of experimental attempts to
demonstrate the existence of such topological surface states
\cite{wol12,bot12,jia13,kim13,gli13}. Yet, ARPES could not
unambiguously reveal \cite{nxu13,neu13,fra13,miy12} the associated
Dirac cones. 
Our focus, however, is not on these topological properties of
SmB$_6$.

In an effort to study the DOS, tunneling \cite{gue82,ams98} and
point contact spectroscopy \cite{fla01,zha13} were conducted, most
of which indicated the opening of a gap of the order of 20 meV.
However, as shown below, locally (atomically) resolved information
on the DOS is required which can be obtained \cite{yee13} by
Scanning Tunneling Microscopy (STM) on SmB$_6$. Furthermore, in
Kondo systems like SmB$_6$ tunneling can take place not only into
the conduction band but also into the 4$f$ states. The
interference of these tunneling processes may give rise to
so-called Fano resonances \cite{fan61,mal09,yan09,fig10,woe10} and
thereby obscure the DOS of interest.

Here we report on the observation of two different types of
surfaces resulting from {\it in situ} low-temperature cleave of
SmB$_6$: large areas of reconstructed surfaces and smaller
(typically some nm, rarely a few ten nm) atomically flat terraces
of varying size and termination. Tunneling spectroscopy on the
latter, non-reconstructed surfaces appears to reveal the DOS of
the material within areas of a few nm while the quantum-mechanical
interference on areas of some ten nm yields excellent agreement
with the prediction for Fano resonances. Clear indications for
hybridization of the coherent 4$f$ and the conduction band as well
as the Kondo effect are derived in both cases. Furthermore, a
finite surface conductance is observed in all cases.

\begin{figure}[t]
\includegraphics[width=8.8cm,clip=true]{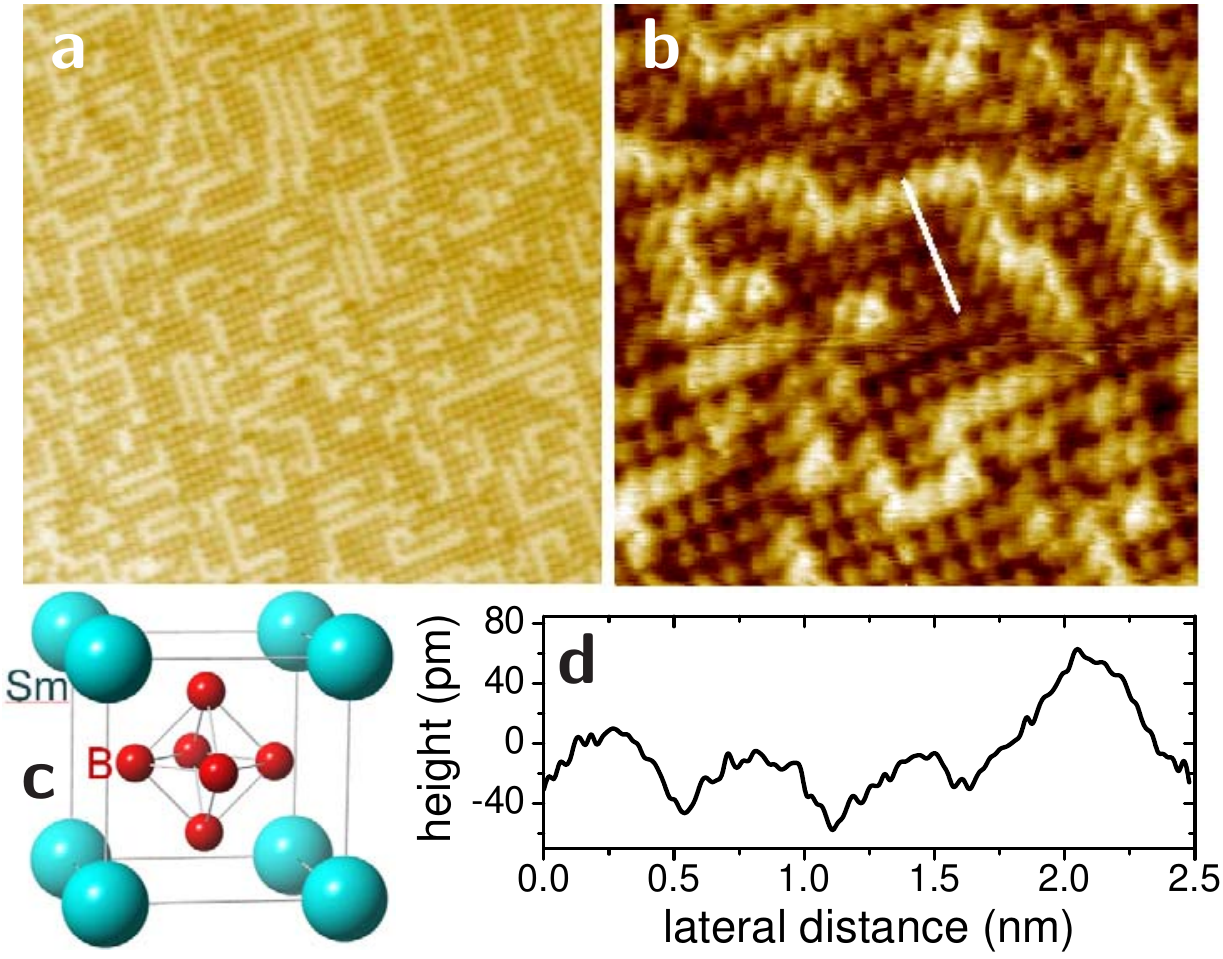}
\caption{a) Topography overview on a reconstructed surface of
SmB$_6$, scan area ($30 \times 30$) nm$^2$, $V =$ 0.2 V, current
set point $I_{sp} =$ 0.6 nA. b) Detailed view of a reconstructed
surface obtained on a different sample, area ($10 \times 10$)
nm$^2$. c) Crystal structure of SmB$_6$. d) Height scan along the
line indicated in b).} \label{recon}
\end{figure}

Single crystals of SmB$_6$ were grown using the aluminum flux
method. SmB$_6$ crystallizes in a CaB$_6$ structure type with
space group $Pm\bar{3}m$ (221), lattice constant $a =$ 4.133
{\AA}, see Fig.\ \ref{recon}(c). We report results obtained on
nine samples, cleaved {\it in situ} at $T \sim 20$ K approximately
along one of the principal cubic crystallographic axes. STM was
conducted in an ultra-high vacuum system (Omicron Nanotechnology)
at $p \lesssim 2 \!\times\! 10^{-9}$ Pa.

The majority of the surfaces (we estimate more than 90\% of the
surface areas investigated so far) exhibited topographies as
exemplified in Figs.\ \ref{recon}(a) and (b). Such topographies
may extend over several $\mu$m and clearly attest the simplest
solution to avoid the polarization catastrophy associated with
polar surface termination planes \cite{nog00}. The presence of Sm
and B$_6$ octahedra in equal amounts and their mixing on short
length scales prevents the build-up of long-range electric fields
from charged surface areas \cite{dam13}. Based on the height
difference between the two types of corrugations, Fig.\
\ref{recon}(d), we interpret the bright lines in Figs.\
\ref{recon}(a) and (b) as chains of Sm atoms residing on top of
otherwise flat B terraces, in good agreement with the proposal of
a partly occupied Sm surface layer \cite{aon79}. Apparently,
depending on the cleave conditions, the Sm and B$_6$ entities can
be disordered (see Fig.\ \ref{recon} and Ref.\ \onlinecite{fra13})
or ordered (see supplemental material II, Fig.\ S2, and Ref.\
\onlinecite{yee13}). It is likely that the surface
reconstructions, both ordered and disordered, heavily influence
the electronic properties of these surfaces. We therefore focus
our spectroscopy work on the non-reconstructed surfaces in the
following.

\begin{figure}[t]
\includegraphics[width=8.8cm,clip=true]{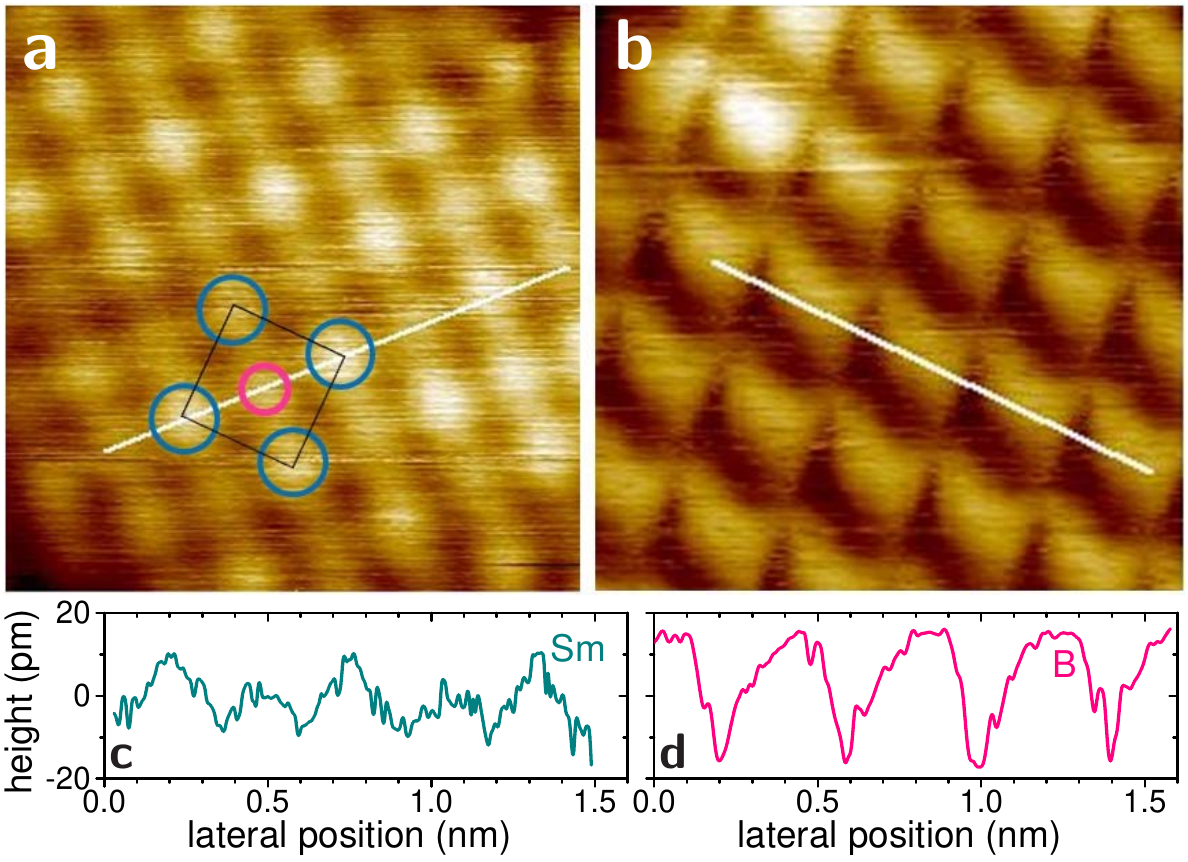}
\caption{Topography on small ($2 \times 2$ nm$^2$)
non-reconstructed surface areas of a) Sm-termination ($V = 0.2$ V,
$I_{sp} =$ 0.6 nA) and b) B termination. Cyan circles in a)
indicate Sm atoms, the pink one encircles a B atom, {\it cf.}
Fig.\ \ref{recon}. c) and d) Height scans along the white line
(parallel to $\langle$110$\rangle$) indicated in a) and the line
parallel to $\langle$100$\rangle$ in b), respectively.}
\label{unrec}
\end{figure}

Only occasionally a second type of topography was observed which
reflects patches of non-reconstructed surface areas. Assuming
cleavage along the $\{$001$\}$ plane through breaking
inter-octahedral B--B bonds \cite{mon01}, Fig.\ \ref{unrec}(a)
exhibits a Sm-terminated surface. This termination is apparent
when considering that the white line runs along a
$\langle$110$\rangle$ direction; the corresponding height scan is
presented in Fig.\ \ref{unrec}(c): The more prominent protrusions
indicate Sm atoms [marked by cyan circles in Fig.\ \ref{unrec}(a)]
spaced about $a \sqrt{2}$ apart, the less pronounced protrusions
visualize the top-most atom of the B octahedra centered between
the Sm atoms along the $\langle$110$\rangle$ diagonal [magenta
circle in Fig.\ \ref{unrec}(a)]. In contrast, Fig.\ \ref{unrec}(b)
shows a different area, with the white line pointing along a
$\langle$100$\rangle$ direction. The periodicity of the
corrugations nicely agrees with the lattice constant $a$ and no
intermediate features are observed. Such surfaces are assigned as
B-terminated. These assignments, and specifically the appearance
of the B octahedra apex atoms between the Sm atoms, are
corroborated by observations on step edges (supplemental material
III). We note that---with only a very few exceptions discussed
below---these non-reconstructed areas are only a few nm in extend.
For two of the investigated samples we did not succeed in finding
any non-reconstructed surface areas at all.

\begin{figure}[t]
\includegraphics[width=8.8cm,clip=true]{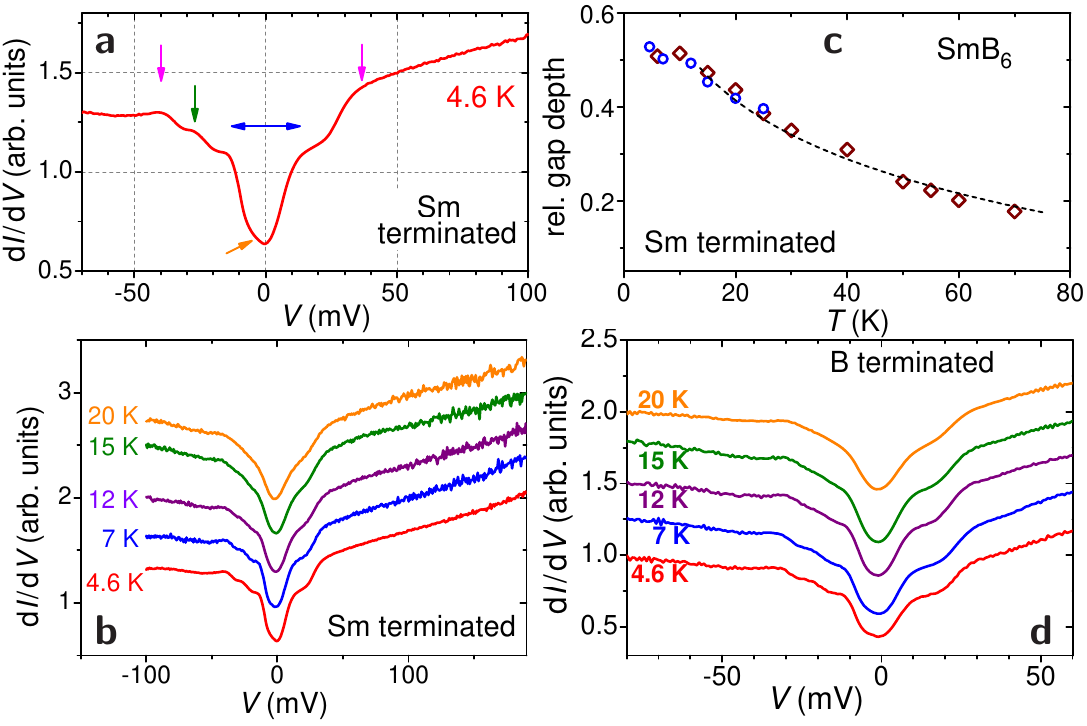}
\caption{a) Tunneling spectroscopy on a Sm-terminated surface at
4.6 K. Features marked by arrows are discussed in the text. b)
Temperature dependent STS on a Sm-terminated surface. ${\rm d}I /
{\rm d}V$-curves for $T \geq 7$ K are offset for clarity. c)
Temperature dependence of the zero-bias dip measured on
Sm-terminated surfaces of two different samples. The dashed line
indicates a logarithmic $T$-dependence. d) $T$-dependent STS on a
B-terminated surface. All spectra are averaged over areas of some
nm$^2$.} \label{spec}
\end{figure}

Scanning tunneling spectroscopy (STS) on the above
non-reconstructed surfaces is presented in Fig.\ \ref{spec}. The
exemplary tunneling conductance $g(V) = {\rm d}I(V) / {\rm d}V$ of
Fig.\ \ref{spec}(a) obtained on a Sm-terminated surface at 4.6 K
reveals several features: i) A partial gap of $\sim$16 meV
develops around the Fermi energy $E_{\rm F}$. This is consistent
with the hybridization gap observed in many other measurements
\cite{zha13,nxu13,neu13,fra13,yee13,gor99}. ii) There is a finite
${\rm d}I / {\rm d}V$-value at $V \! = \! 0$, {\it i.e.} a finite
DOS at $E_{\rm F}$. This might indicate an incomplete (or pseudo-)
gap, but is also compatible with an additional conductance channel
at the surface. iii) Peaks are observed at around $\pm 40$ meV.
These excitations likely result from a $\mathcal{J} \! = \! 0$ to
$\mathcal{J}\! = \! 1$ inter-multiplet transition of the Sm$^{2+}$
ion as seen in neutron scattering experiments \cite{ale93}. iv) A
small peak at around $-27$ meV might be related to crystalline
electric field (CEF) excitations \cite{ant02} between the
$\Gamma_8$ quartet and the $\Gamma_7$ doublet of Sm$^{3+}$, yet no
such indications were found in neutron scattering experiments
\cite{ale00b}. If we assume a fully developed bulk hybridization
gap at low $T$, then the CEF excitations should take place out of
states close to the gap edges yielding a CEF excitation energy of
11 meV in line with reported values \cite{nyh95}. v) There seems
to be an additional hump close to $-3$ meV, {\it i.e.} inside the
hybridization gap. One may speculate that this feature is related
to in-gap states seen in other measurements
\cite{fla01,miy12,gor99,gab01,noz02,cal07}. Specifically, it is
only observed for $T \leq 12$ K, Fig.\ \ref{spec}(b), as seen in
many of those measurements.

The temperature evolution of these features on a Sm-terminated
surface can be inferred from Fig.\ \ref{spec}(b). At 20 K, the
partial hybridization gap and the inter-multiplet transitions,
features i) and iii), can barely be disentangled. To investigate
the hybridization gap further, we studied the $T$-dependence of
its depth at $V \! = \! 0$ relative to the tunneling conductance
at $|V| \! \geq \! 50$ mV (see supplemental material IV). The
$T$-dependent closing-up of this partial gap for two different
samples is presented in Fig.\ \ref{spec}(c). The experimental
values for the normalized depth of this gap agrees well with
results for the single-ion conductance within numerical
renormalization group (NRG) calculations \cite{cos00} which
predict a logarithmic decay, dashed line in Fig.\ \ref{spec}(c).
Such a logarithmic $T$-dependence is a hallmark of Kondo physics
and strongly supports a scenario based on the Kondo effect
\cite{aep92,ern11}. The fit (dashed line) yields a Kondo
temperature $T_{\rm K} \approx 39$ K, in line with other estimates
for the opening of the hybridization gap. Deviations from this
logarithmic $T$-dependence due to departure from single-ion
behavior, {\it i.e.} due to enhanced lattice effects of the Kondo
ions, become apparent at $T \!\lesssim\! 15$ K, in a regime where
also possible in-gap states appear (feature v).

Temperature dependent ${\rm d}I / {\rm d}V$-curves obtained on an
non-reconstructed B-terminated surface are presented in Fig.\
\ref{spec}(d). As expected, the hybridization gap exhibits a very
\begin{figure}[t]
\includegraphics[width=8.8cm,clip=true]{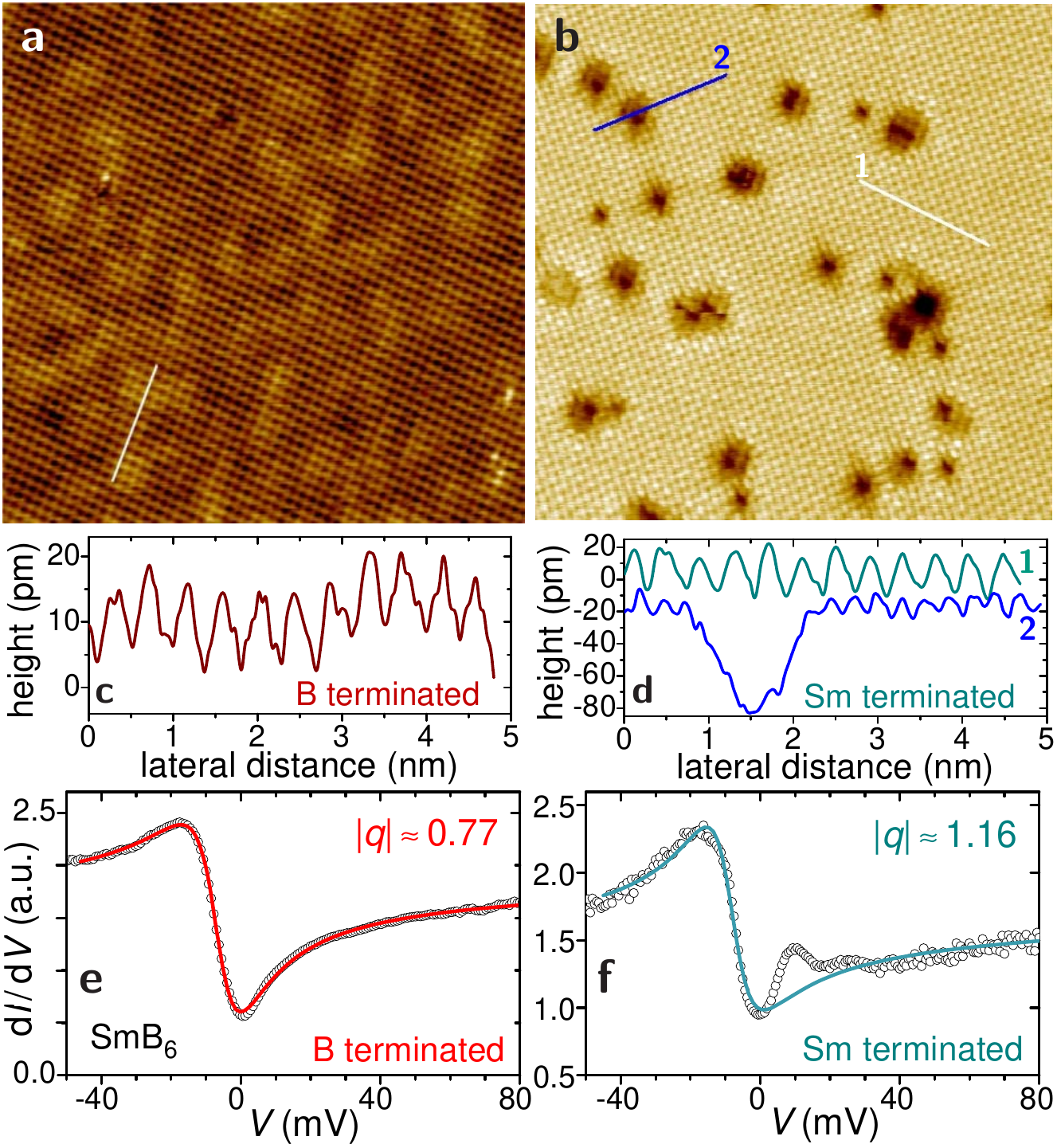}
\caption{a) Comparison between larger, non-reconstructed
B-terminated (left) and Sm-terminated surfaces (right). a) and b)
Topographies on B- ($20 \times 20$ nm$^2$) and Sm-surfaces ($17.2
\times 17.2$ nm$^2$), respectively . c) and d) Height scans along
the lines marked in a) and b). White lines are parallel to
$\langle$100$\rangle$ directions, blue line 2 in b) and d) is
along $\langle$110$\rangle$. e) and f) Tunneling spectroscopies
(circles) at 4.6 K on B- and Sm-terminated surfaces, respectively.
The lines are fits to the Fano formula; the obtained values of $q$
are indicated.} \label{fano}
\end{figure}
similar behavior as on Sm-terminated surfaces. However, all other
features appear to be shifted towards smaller absolute values of
energy (or $|V|$). This highlights the necessity to involve
spectroscopic techniques that allow to discriminate between
differently terminated surfaces when analyzing SmB$_6$ surfaces.
However, also the B-terminated surfaces exhibit a finite
conductance at $V \! = \! 0$ again compatible with an additional
(surface) conductance channel.

Only very rarely did we succeed in locating larger patches of
non-reconstructed areas which clearly exceed lateral dimensions of
10 nm$^2$ (so far, such patches were found only on two cleaved
surfaces). Two such examples are shown in Figs.\ \ref{fano}(a) and
(b). The height scans along the white lines in (a) and (b), which
are presented in Figs.\ \ref{fano}(c) and (d) respectively, were
conducted parallel to $\langle$100$\rangle$ directions and confirm
the expected unit-cell spacing $a$ between the corrugations. There
are no discernible intermediate corrugations at any
$\langle$110$\rangle$ direction in Fig.\ \ref{fano}(a). However,
they are obvious in (b): The height scan along the $[$110$]$
direction, blue lines marked 2 in Figs.\ \ref{fano}(b) and (d),
exhibits corrugations spaced by about $a/\sqrt{2}$ as in Fig.\
\ref{unrec}(a). Consequently, we attribute the surface of Fig.\
\ref{fano}(a) to a B-terminated one, while Fig.\ \ref{fano}(b)
represents a Sm-terminated surface. The latter surface exhibits
dents of about 80 pm in depth, see blue line 2 in Fig.\
\ref{fano}(d), corresponding to the height difference between Sm
and the top-most B atom on a Sm-terminated surface ({\it cf.}
supplemental material III). This suggests that these dents are
caused by missing Sm atoms probably ripped out of the Sm surface
layer while cleaving the sample. In contrast, there are only small
height variations of some pm resulting in patches of a few lattice
constants in extent.

Conducting STS on such an extended non-reconstructed B surface
provided ${\rm d}I / {\rm d}V$-curves as exemplary shown in Fig.\
\ref{fano}(e) for $T = 4.6$ K. Clearly, such a tunneling
conductance is indicative of a Fano resonance at energy $E_0$
which results from tunneling into {\it two} coupled channels
\cite{fan61}, namely the conduction band and the quasiparticle
states \cite{mal09,fig10,woe10}. Apparently, these
non-reconstructed areas are large enough to establish coherence in
the two tunneling channels. The resulting tunneling conductance
can then be expressed as \cite{sch00}:
\begin{equation}
g(V) \propto \frac{(\epsilon + q)^2}{\epsilon^2 + 1} \; , \qquad
\epsilon = \frac{2(eV - E_0)}{\Gamma} \: . \label{fanf}
\end{equation}
Here, the asymmetry parameter $q$ is related to the ratio of
probabilities for tunneling into the 4$f$ states {\it vs.} into
the conduction band while $\Gamma$ describes the resonance width.
Eq.\ (\ref{fanf}) was successfully applied to single-impurity
\cite{mad98,li98} as well as Kondo lattice systems
\cite{sch10,ayn12}. Clearly, eq.\ (\ref{fanf}) also describes our
data excellently even at low $T$ and without considering details
of the DOS \cite{zha13}. The corresponding fit, red line in Fig.\
\ref{fano}(e), yields $\Gamma = 16.5$ mV, a value which agrees
well with the width of the hybridization gap discussed above. As
to be expected for a B-terminated surface, the value of $|q| =
0.77$ is smaller than unity consistent with predominant tunneling
into the conduction band. Note that the asymmetry of the ${\rm
d}I(V) / {\rm d}V$-curves (and hence the sign of $q$) is a direct
consequence of the particle-hole asymmetry of the conduction band
\cite{fig10}; the asymmetry shown here is consistent with other
STM measurements on a reconstructed surface \cite{yee13} but
opposite to results from point-contact spectroscopy \cite{zha13}.

An overall very similar behavior is observed for spectra obtained
on a Sm-terminated surface, Fig.\ \ref{fano}(f), with the
exception of two additional excitations at around 10 and 25 meV.
If these two excitations are ignored, again, a good fit of eq.\
(\ref{fanf}) to the data is possible, line in Fig.\ \ref{fano}(f).
The so obtained value $\Gamma = 16.4$ meV agrees with that of the
B-terminated surface, as expected for the hybridization. However,
$|q| = 1.16$ indicates a more pronounced tunneling into the 4$f$
quasiparticle states which appears reasonable on a Sm-terminated
surface. Both additional excitations have also been seen in
earlier measurements in the tunneling regime \cite{fla01}. One may
speculate that the excitation at 25 meV has its counterpart at
$-27$ meV at smaller non-reconstructed Sm surfaces, feature iv)
above, with the respective features at opposite voltage signs
masked by other, more pronounced excitations.

Also the larger non-reconstructed areas exhibited---as all the
investigated surfaces---a finite conductance at $V \! = \! 0$.
Consequently, our locally resolved measurements show that the
surface reconstruction alone cannot account for the surface
conductance \cite{dam13}. The robustness of the measured
conductance at $V \! = \! 0$ is consistent with an additional
conductance channel, possibly at the surface.

In conclusion, we observed two different types of surfaces:
predominantly reconstructed surfaces, but occasionally also
non-reconstructed ones. For the latter, Sm- as well as
B-terminations were found, but larger areas (lateral extension
well beyond 10 nm) were only encountered very rarely. STS on
non-reconstructed surfaces indicated the formation of a
hybridization gap with logarithmic $T$-dependence and $T_{\rm K}
\approx 40$ K as well as inter-multiplet transitions of the
Sm$^{2+}$. Spectra on larger non-reconstructed areas showed
excellent formation of Fano resonances establishing the underlying
hybridization effect. All surfaces exhibited a finite conductance
consistent with SmB$_6$ being a topological Kondo insulator.

\section*{Acknowledgements}
The authors are indebted to J. W. Allen, P. Coleman, A.
Damascelli, C. Geibel, S. Kirchner, Q. Si, P. Thalmeier and S. von
Moln\'{a}r for stimulating discussions. Research at UC Irvine was
supported by NSF-DMR-0801253.

\end{document}